\documentclass[%
superscriptaddress,
 amsmath,amssymb,
aps,
pra,
floatfix,
twocolumn
]{revtex4-2}

\usepackage{graphicx}
\usepackage{dcolumn}
\usepackage{bm}
\usepackage{xcolor}
\usepackage{physics}
\usepackage{qcircuit}
\usepackage{hyperref}

\begin{document}
\title{When Clifford benchmarks are sufficient; estimating application performance with scalable proxy circuits}
\author{Seth Merkel }
\affiliation{IBM Quantum}
\email{seth.merkel@ibm.com}
\author{Timothy Proctor}
\affiliation{Quantum Performance Laboratory, Sandia National Laboratories}
\author{Samuele Ferracin}
\affiliation{IBM Quantum}
\author{Jordan Hines}
\affiliation{Quantum Performance Laboratory, Sandia National Laboratories}
\affiliation{Department of Physics, University of California, Berkeley}
\author{Samantha Barron}
\affiliation{IBM Quantum}
\author{Luke~C.~G.~Govia}
\affiliation{IBM Quantum}
\author{David McKay}
\affiliation{IBM Quantum}
\date{\today}

\begin{abstract}
The goal of benchmarking is to determine how far the output of a noisy system is from its ideal behavior; this becomes exceedingly difficult for large quantum systems where classical simulations become intractable. A common approach is to turn to circuits comprised of elements of the Clifford group (e.g., CZ, CNOT, $\pi$ and $\pi/2$ gates), which probe quantum behavior but are nevertheless efficient to simulate classically. However, there is some concern that these circuits may overlook error sources that impact the larger Hilbert space.  In this manuscript, we show that for a broad class of error models these concerns are unwarranted.  In particular, we show that, for error models that admit noise tailoring by Pauli twirling, the diamond norm and fidelity of any generic circuit is well approximated by the fidelities of proxy circuits composed only of Clifford gates.  We discuss methods for extracting the fidelities of these Clifford proxy circuits in a manner that is robust to errors in state preparation and measurement and demonstrate these methods in simulation and on IBM Quantum's fleet of deployed heron devices.         
\end{abstract}

\maketitle

\section{Introduction}
As quantum processors grow in complexity there have been a number of recent works attempting to demonstrate quantum advantage or utility \cite{QuantumVolume, supremacy19,utility,supremacy24,RCSquantinium}. As more performant quantum simulations are realized, the field of classical simulation of quantum systems has correspondingly advanced (see \cite{expvssim} and the references therein).  This is a relationship that has pushed forward the state-of-the-art in both fields, but presents challenges for the sub-field of benchmarking and characterization.  In order to compare and contrast the performance of larger machines, it is essential that our benchmarks do not require exponentially increasing amounts of classical computation to validate the output from our quantum processor \cite{Proctor2025-cd, Amico2023-ze}.  Ideally, we would benchmark quantum processors using applications that are thought to be hard classically but admit efficient classical verification algorithms, for example factoring large numbers \cite{shor} or other slightly more tractable cryptographic one-way functions \cite{Mahadev}.  Unfortunately such algorithms seem out of reach for the current generation of quantum hardware.  Another option, the one we consider in this work, is to benchmark quantum processors by running quantum computations that are not hard to simulate classically in the hopes that they provide some insights into that processor's performance on other quantum computations that solve classically-difficult problems.  A very common technique is to look at circuits composed of Clifford gates which are known to be classically simulatable due to the Gottesmann-Knill theorem \cite{GKtheorem}.  Some examples are randomized benchmarking and its variants \cite{KnillRB,MagesanRB,HelsenRB,birb,layerfid}, or accreditation \cite{ogaccred,ExpAccred}.

The question we attempt to answer in this manuscript is: When is it possible to accurately predict the performance of general circuits when executing only Clifford circuits?  The answer to this question, at least in part, must depend on the error model being considered.  If non-Clifford gates have far more error than the Clifford ones (e.g., T-gate distillation in the surface code), Clifford gates alone probably aren't sufficient to make more general performance bounds.  Likewise, if the system is strongly non-Markovian one can dream up pathological errors models for which no set of test circuits accurately predicts the performance of future experiments.  In this manuscript we will consider an error model we refer to as the Pauli twirling assumption (PTA), and will show that in this error model Clifford circuits are indeed sufficient to approximately bound the performance of generic circuits. We will describe the PTA in detail in Sec.~\ref{Sec:PTA}, but its essence is that it describes a family of error models for which one can perform Pauli twirling to tailor the effective error channels to be Pauli stochastic \cite{twirl, GellarPTA,RandomCompiling}.  This can be enforced by insisting that the local error processes are independent of the particular choice of single qubit gates, although experimental evidence suggests it is also true under broader conditions \cite{Hashim2023-qk}.  

In Sec.~\ref{Sec:Bounds}, we will show that this independence from the choice of single qubit gates also extends to aggregate properties of the circuits, namely the overall process fidelity and diamond distance, at least to lowest order in the circuit's error rate. If we `Cliffordize' a generic test circuit by replacing the single-qubit operations by random single-qubit Clifford operations (assuming the multi-qubit gates were Clifford to begin with) we obtain a family of efficiently verifiable proxy circuits with approximately the same fidelity and diamond norm as the original.  This is very analogous to the trap circuits bounds from accreditation \cite{ogaccred} or the observations of Clifford benchmarks in \cite{quadLoss}.  In Sec.~\ref{sec:spamrobustdfe} we discuss how to extract the fidelity of Cliffordized circuits using variants of direct fidelity estimation, \cite{directfid, Moussa2012-rq}, that have been made robust to SPAM errors.  Finally, in Sec.~\ref{sec:exp} we corroborate some of the arguments in this manuscript with experimental and numerical data, we construct a large scale volumetric benchmark, \cite{volumetric}, which we run on the IBM deployed devices {\it ibm\_fez, ibm\_kingston, ibm\_marrakesh} and {\it ibm\_torino} and compare to the estimates from random circuit sampling \cite{supremacy19,supremacy24}.

\section{Pauli twirling assumption}\label{Sec:PTA}
\begin{figure*}
\centering
\begin{align}
\begin{array}{c}
\Qcircuit @C=1em @R=1em {
     \lstick{\ket{0}^{S_1}} & \gate{U}  & \multigate{5}{{\mathcal E}_{\rm prep}}&
        \gate{U}  & \multigate{5}{{\mathcal C}_1} & \multigate{5}{{\mathcal E}_1}&
        \gate{U}  & \multigate{5}{{\mathcal C}_2} &  \multigate{5}{{\mathcal E}_2}& \qw  & \cdots&&
        \gate{U}  & \multigate{5}{{\mathcal C}_L} &  \multigate{5}{{\mathcal E}_L}&
        \gate{U} &  \multigate{5}{{\mathcal E}_{\rm meas}}&
        \gate{U}&\measureD{Z^{M_1}}\\
     \lstick{\ket{0}^{S_2}}& \gate{U}  & \ghost{{\mathcal E}_{\rm prep}}&
        \gate{U}  & \ghost{{\mathcal C}_1} &  \ghost{{\mathcal E}_1} &
        \gate{U}  & \ghost{{\mathcal C}_2} &  \ghost{{\mathcal E}_2} & \qw&\cdots &&
        \gate{U}  & \ghost{{\mathcal C}_L} &  \ghost{{\mathcal E}_L}&
        \gate{U}& \ghost{{\mathcal E}_{\rm meas}}&
        \gate{U}&\measureD{Z^{M_2}}\\
    \lstick{\ket{0}^{S_3}} & \gate{U}  & \ghost{{\mathcal E}_{\rm prep}}&
        \gate{U} &\ghost{{\mathcal C}_1}&  \ghost{{\mathcal E}_1}&
            \gate{U} &\ghost{{\mathcal C}_2}&  \ghost{{\mathcal E}_2}&  \qw  &\cdots&&
        \gate{U}  & \ghost{{\mathcal C}_L} &  \ghost{{\mathcal E}_L}&
        \gate{U}&\ghost{{\mathcal E}_{\rm meas}}&
        \gate{U}&\measureD{Z^{M_3}}\\
     \lstick{\ket{0}^{S_4}}& \gate{U}  & \ghost{{\mathcal E}_{\rm prep}}&
        \gate{U} &\ghost{{\mathcal C}_1}&  \ghost{{\mathcal E}_1}&
            \gate{U} &\ghost{{\mathcal C}_2}&  \ghost{{\mathcal E}_2}&  \qw  &\cdots&&
        \gate{U}  & \ghost{{\mathcal C}_L} &  \ghost{{\mathcal E}_L}&
        \gate{U}&\ghost{{\mathcal E}_{\rm meas}}&
        \gate{U}&\measureD{Z^{M_4}}\\
    \lstick{\ket{0}^{S_5}} & \gate{U}  & \ghost{{\mathcal E}_{\rm prep}}&
        \gate{U} &\ghost{{\mathcal C}_1}&  \ghost{{\mathcal E}_1}&
            \gate{U} &\ghost{{\mathcal C}_2}&  \ghost{{\mathcal E}_2}&  \qw  &\cdots&&
        \gate{U}  & \ghost{{\mathcal C}_L} &  \ghost{{\mathcal E}_L}&
        \gate{U}&\ghost{{\mathcal E}_{\rm meas}}&
        \gate{U}&\measureD{Z^{M_5}}\\
    \lstick{\ket{0}^{S_6}} & \gate{U}  & \ghost{{\mathcal E}_{\rm prep}}&
        \gate{U} &\ghost{{\mathcal C}_1}&  \ghost{{\mathcal E}_1} &
            \gate{U} &\ghost{{\mathcal C}_2}&  \ghost{{\mathcal E}_2}&  \qw  &\cdots&&
        \gate{U}  & \ghost{{\mathcal C}_L} &  \ghost{{\mathcal E}_L}&
        \gate{U}&\ghost{{\mathcal E}_{\rm meas}}&
        \gate{U}&\measureD{Z^{M_6}}
        }
\end{array}\nonumber
\end{align}

\caption{General layered circuit with prep and measurement.  Here each $U$ corresponds to an arbitrary single qubit gate.  The ${\mathcal C}$ terms denote multiqubit Clifford layers and the ${\mathcal E}$  denote error processes.  We may marginalize over some subset of the measurement outcomes, and correspondingly may only be sensitive to some subset of the qubit initializations, denoted by the binary vectors $M_j$ and $S_j$ respectively.  }    \label{fig:general_circuit} 
\end{figure*}

As stated in the introduction, it is impossible to make inferences about performance without some assumptions on error models.  In this manuscript, we look at a family of error models that admit Pauli twirling, which we refer to as the Pauli twirling assumption or PTA.  Error models like the PTA have been considered in the context of randomized compiling \cite{RandomCompiling} as well as accreditation \cite{ExpAccred}, but for completeness we will provide our own pedantic definition.  It should be noted that these error models are defined out of mathematical convenience and not derived from well-posed microscopic device models.  Nevertheless, Pauli twirling has been shown to work fairly well on superconducting systems, \cite{Hashim2023-qk, RCExp}, and at least anecdotally on other physical platforms as well.

To begin, we restrict our circuits under consideration to those where all multi-qubit operations are elements of the Clifford group, and therefore all non-Clifford gates are single qubit operations. Two standard examples are the Clifford$+$T gateset and CNOTs with arbitrary single qubit gates. In this formalism, any computation can be implemented with a circuit that consists of alternating layers of generic single qubit gates and multi-qubit Clifford layers, and we assume circuits of this structure. It is natural then to describe the error on a particular pair of layers as depicted in Fig.~\ref{fig:general_circuit}, that is we imagine the noise as an operation inserted after every multi-qubit Clifford layer (Fig.~\ref{fig:general_circuit} also includes state preparation and measurement error, or SPAM, but we will address that later).  With the notation from Fig.~\ref{fig:general_circuit} the formal statement of the PTA is as follows:

\begin{center}
\fbox{\begin{minipage}{20em}
\begin{center}
{\bf Pauli Twirling Assumption}
\end{center}
The layer errors for an $n$-qubit circuit, $\{ {\mathcal E}_1,{\mathcal E}_2,\ldots {\mathcal E}_L \}$, for a fixed choice of multi-qubit gate layers, $\{ {\mathcal C}_1,{\mathcal C}_2,\ldots {\mathcal C}_L \}$, can be described by fixed $n$-qubit process matrices that are independent of the choice of single qubit gates $U$.
\end{minipage}}
\end{center}

The independence of the error process from the single qubit gate layers allow us to consider a large family of circuits that have exactly the same error characteristics, in the sense that they have the same error maps on their layers.  In fact, some circuits not only have the same error characteristics but also describe the same logical operation, such as the family of circuits obtained by Pauli twirling \cite{GellarPTA,RandomCompiling}.  When we average over these Pauli twirls, the effective layer errors, ${\mathcal E}_i$, can be described as Pauli stochastic channels, that is a convex combination ${\cal E}_i(\rho) = \sum_j p_{i,j} P_j \rho P_j$.  Alternatively, the effect of a Pauli twirl in the Pauli transfer matrix (PTM) representation of ${\cal E}_i$ is to set all off-diagonal elements of the PTM to zero.    

It is worth noting that our definition of the PTA is not strictly Markovian.  If we repeat the same layer of multi-qubit gates many times in our circuit we do not assume that the error process is identical for each application.  That is, we can still effectively Pauli twirl if some filter or heating process changes the error terms for subsequent applications of the layer.  We do, however, assume Markovianity in the sense that the error processes are the same shot to shot when we repeat the total circuit many times.  In this ``shot to shot'' Markovian model, Pauli channel learning methods that rely on layer amplification (i.e., cycle benchmarking \cite{cycle} or PEC learning \cite{PauliLearning}) are inapplicable; one needs to explore the error processes in situ.  

For the remainder of this manuscript we will consider families of circuits on $n$ qubits with $L$ layers where we have fixed the multi-qubit layers and thus the error channels.  Furthermore, we assume the errors have been tailored with Pauli twirling.  Elements of this set of circuits can be fully described by a ${3n(L+1)}$ real parameters (Euler angles for each of the single qubit operations), and we will denote the circuit's ideal operation by the channel ${\mathcal U}$, which is a unitary.  The goal of this paper will be to show that all of the salient error properties of this family of circuits can be obtained by looking at a subset, ${\mathcal C}$, that restricts the single qubit operations to be elements of the Clifford group, a efficiently simulatable subset of circuits.

\section{Bounding performance in the PTA}\label{Sec:Bounds}
In this section we will show that, in the PTA, the diamond distance error between the noisy and ideal implementation of any circuit (denoted $ d_\diamond({\mathcal U})$) is equal to the infidelity of any Cliffordization of that circuit ($ r(\mathcal{C})$), up to higher-order terms in the infidelity:
\begin{align}
    d_\diamond({\mathcal U}) = r(\mathcal{C}) + {\mathcal O}( r(\mathcal{C})^2).
\end{align}
That is, we can approximate the diamond distance of any circuit from its ideal implementation,
\begin{align}
    d_\diamond({\mathcal U}) &= \frac{\|{\mathcal U}^{\rm ideal} - {\mathcal U}^{\rm noisy} \|_\diamond}{2}\nonumber\\
    &= \frac{1}{2} \max_\rho \|({\mathcal U}^{\rm ideal} - {\mathcal U}^{\rm noisy})\otimes{\mathbb I}^{2^n} (\rho)  \|_1,
\end{align}
 by the process fidelity of a much simpler to validate Clifford circuit,
 \begin{align}
    r({\mathcal C}) &= 1-{\rm Tr}\left({\rm PTM}({\mathcal C}^{\rm ideal})^T {\rm PTM}({\mathcal C}^{\rm noisy}) \right)/4^n \nonumber\\
    &=\frac{2^n}{2^n+1} \left(1 - \int {\rm d} \vert \psi \rangle F\left({\mathcal C}^{\rm ideal} (\vert \psi \rangle ),{\mathcal C}^{\rm noisy} (\vert \psi \rangle)\right)\right).
\end{align}
Here we have used a superscript on ${\mathcal U}$ to denote the ideal and noisy implementation and will reserve subscripts to describe the layers, that is ${\mathcal U} = {\mathcal U}_L \ldots {\mathcal U}_2 {\mathcal U}_1$.  We've also implicitly assumed that the ideal operation is unitary.  

There are essentially two steps to the argument,
\begin{enumerate} 
    \item All circuits in the PTA with fixed multi-qubit layers have approximately the same process infidelities.
    \item All circuits in the PTA with fixed multi-qubit layers have diamond distances that are approximately equal to their process infidelity. 
\end{enumerate}
When taken together, this allows us to estimate the diamond norm (or process fidelity) for any circuit by the process fidelity of any other proxy circuit.  We will choose these proxy circuits to be composed solely of Clifford gates. 
 Numerical and experimental evidence to support these claims are shown in Sec.~\ref{Sec:fidsthesame}. 

\subsection{Infidelity is almost the same for all circuits in the PTA}
We can do almost all of the heavy lifting in the this subsection by the work in \cite{polardecomp,trustinqc}.  Let's begin by defining
\begin{align}
    \bar{r} \equiv \sum_{j=1}^L r({\mathcal U}_j).
\end{align}
This is slightly different notation from that of the previous two references where they typically refer the average instead of the sum.  Crucially, $\bar{r}$ has no dependence on ${\mathcal U}$ since the layer by layer error terms have no dependence on the choice of single qubit gates in the PTA.  The essential observation drawn from \cite{polardecomp,trustinqc} is that the process fidelity, $1-r$, of a circuit in the PTA is approximately the product of the process fidelities of the individual layers,
\begin{align}
    1-r({\mathcal U}) = \prod_{j=1}^L \left(1-r({\mathcal U}_j) \right) + {\mathcal O} (\bar{r}^2),
\end{align}
where here we have assumed that $n \gg 1$ so that we can ignore $\mathcal{O}(1/4^n)$ factors (it is ``process polarizations'' \cite{Hashim2024-om} that approximately multiply, which are an $n$-dependent rescaling of process fidelities that differ from process fidelities by an  $\mathcal{O}(1/4^n)$ factor).
We can shuffle the terms around to see that
\begin{align}
    r({\mathcal U}) = \bar{r} + {\mathcal O} (\bar{r}^2) \label{eq:rfromrbar},
\end{align}
collecting terms of the form $r({\mathcal U}_j) r({\mathcal U}_k)$ into the ${\mathcal O} (\bar{r}^2)$ corrections. Therefore, all circuits with fixed multi-qubit gates have equivalent infidelities, with corrections to order $\bar{r}^2$.  In particular, for any test circuit, ${\mathcal U}$, and one of its Cliffordizations, ${\mathcal C}$, 
\begin{align}
    r(\mathcal{U}) = r(\mathcal{C}) + {\mathcal O}( r(\mathcal{C})^2).
\end{align}
This means that we can estimate $ r(\mathcal{U})$ by instead measuring $r(\mathcal{C})$.

\subsection{Diamond distance is approximately the infidelity in the PTA}
 We will now show that $\bar{r} + {\mathcal O}(\bar{r}^2) \le d_\diamond({\mathcal U}) \le \bar{r}$.
When combined with the above results, this then implies that
\begin{equation}
    d_\diamond({\mathcal U}) =  r(\mathcal{C}) + {\mathcal O}( r(\mathcal{C})^2). \label{eq:d=r+r2}
\end{equation}
Both the upper and lower bounds are fairly straightforward.

For the lower bound we have that
\begin{align}
    d_\diamond({\mathcal U}) &\ge r({\mathcal U})=\bar{r}+{\mathcal O}(\bar{r}^2).
\end{align}
This is because the diamond distance always upper bounds process infidelity, \cite{Rbconfidence}, which we can then express in the form from Eq.~\ref{eq:rfromrbar}. For the upper bound we have
\begin{align}
    d_\diamond({\mathcal U}) &\le \sum_{j=1}^L d_\diamond({\mathcal U}_j) = \sum_{j=1}^L r({\mathcal U}_j) = \bar{r}
\end{align}
The first inequality is a well-known property of the diamond norm that is shown in \cite{aharanovkitaev}.  Since the layer errors are Pauli stochastic we have that $d_\diamond({\mathcal U}_j) = r({\mathcal U}_j)$, \cite{MagesanRB}.  With both bounds together we see that the diamond distance is approximately equal to $\bar{r}$, the infidelity.  Any corrections are at most ${\mathcal O}(\bar{r}^2)$.    

The results of this section show that we can obtain approximately estimate the diamond norm of any circuit by the process fidelity of any other proxy circuit with the same pattern of two qubit gates.  As we will show in the next section, Clifford circuits are a good choice for these proxy circuits due to the ease of measuring their process fidelities.  

\section{Estimating fidelities in a SPAM robust manner}\label{sec:spamrobustdfe}

We have shown that, in the PTA, the infidelity of any circuit (e.g., a Clifford circuit) approximates the diamond distance for any other circuit with the same pattern of multi-qubit gates.  What remains is to present a method for extracting the infidelity of an arbitrary Clifford circuit.  Our primary tool for this will be a variant of direct fidelity estimation \cite{directfid, Moussa2012-rq} inspired by the techniques developed for binary RB (BiRB) \cite{birb}.

The process fidelity of an error channel, ${\mathcal E}$, is given by the mean of the diagonal elements of the PTM, $\frac{1}{|{\mathcal P}|}\sum_{P_i \in {\mathcal P}}\textrm{Tr}(P_i {\mathcal E} [P_i])$.  If the ideal channel is the identity we can extract these diagonal elements by randomly sampling an element of the Pauli group, preparing one of its $+1$, separable eigenstates as input and measuring the Pauli observable's expectation value at the output. Averaging over different Pauli observables  yields the process fidelity, in the Monte Carlo sense.  When the circuit under interrogation is a more general Clifford circuit, we modify the protocol to do the following: again choose a Pauli uniformly at random from the $n$-qubit Pauli which we measure at the output, but now backwards propagate the Pauli through the Clifford circuit and prepare a $+1$, separable eigenstate of this backwards propagated observable. Because of the PTA, any single qubit gates used for state preparation and measurement, of these Pauli eigenstates and observables, can be absorbed into the first and last layers of the layered circuit without changing the errors. The only additional error sources added to the circuit from this protocol are from preparing the computational $\vert 0\rangle$ state and measuring $Z$'s on all the qubits, that is, SPAM error. These errors can also mostly be Pauli twirled, as discussed in \cite{ExpAccred} and so we will assume that they are described by Pauli stochastic channels and therefore also contribute to the fidelity in a roughly multiplicative manner.

In the remainder of this section we will
present two methods for making direct fidelity estimation SPAM robust, enabling reliable estimating $r(\cal C)$ for an arbitrary Clifford circuit $\cal C$. These methods are presented in order of increasing ease of implementation, but also increasing demands on error assumptions beyond the PTA.  Comparisons on an IBM deployed device are discussed in Sec.~\ref{sec:SPAM_test}.

\subsection{SPAM removal with a reference measurement}\label{Sec:ref_measurement}

Broadly speaking, there have been two main approaches to making fidelity estimation protocols robust.  The first is to look at sequences of varying lengths and fit the decay to an exponential model as done in randomized benchmarking \cite{KnillRB,MagesanRB} or BiRB \cite{birb}.  Since we are looking at a specific circuit and Cliffordization of that circuit, the circuits have fixed length and there is no decay to fit.  The second approach, and our starting point in this manuscript, comes from extensions of mirror benchmarking that enable estimating circuit process fidelities, \cite{trustinqc}, and use the fact that the fidelities are roughly multiplicative to divide out the SPAM error using reference circuits.     

There are a couple issues extending the mirror benchmarking techniques directly to our work.  The first problem is that the effect of SPAM errors on direct fidelity estimation is not independent of the choice of single qubit gates.  The preparation and measurement observables are dependent on the choice of single qubit gates.  The resulting observables likely have support over only a subset of qubits (here support is defined as the qubits for which the Pauli term is not $I$). The SPAM errors on qubits outside of the support are essentially irrelevant, so the expectation value is insensitive to local SPAM errors on those qubits, which implies the choice of single qubit gates affect the magnitude of the SPAM errors in a very dramatic way.  Since we are averaging over Pauli observables this alone is not a huge issue, but the other tricky consideration is that our circuits are not all logically equivalent to the identity, whereas mirror circuits (both the test and reference mirror circuits) are. With mirror circuits, the input and output support are therefore perfectly correlated for both the test circuit and the SPAM reference circuit,  but this is not the case for the more general Clifford circuits we consider in this manuscript.

To address these issues we will ``scramble'' the support with a short layered circuit, ${\mathcal L}$, sampled from some appropriate ensemble of layered Clifford circuits. The idea is illustrated in Fig.~\ref{fig:spamreference}.  We have added a scrambling circuit to both the Cliffordized proxy circuit and the SPAM-reference.  This randomization step ensures that the input and output Paulis are effectively uncorrelated in both cases.  When we divide the fidelity of the Cliffordized circuit by that of the reference, we effectively remove contributions to the error both from the SPAM but also from the log-depth circuit, ${\mathcal L}$, leading to a direct fidelity estimate of ${\mathcal C}$ alone.     

\begin{figure}[h]
\centering
\begin{align}
\begin{array}{c}
{\rm Target~circuit} \\
\Qcircuit @C=1em @R=1em {
         & /^{n} \qw &  \gate{\cal U}  & \qw
       }\\\\
{\rm Cliffordized~circuit} \\
\Qcircuit @C=1em @R=1em {
        \lstick{\ket{0^{\otimes S}}} & /^{n} \qw & 
        \gate{\cal C}& 
        \gate{\cal L}  &\measureD{Z^{\otimes M}}
       }\\\\
{\rm SPAM~reference~circuit}\\
\Qcircuit @C=1em @R=1em {
        \lstick{\ket{0^{\otimes S}}} & /^{n} \qw & 
        \gate{\cal L}  &\measureD{Z^{\otimes M}}
       }\\\\
\end{array}\nonumber
\end{align}

\caption{Removing SPAM with a scrambled reference experiment.  If we divide the fidelity of the Cliffordized circuit by that of the SPAM reference circuit, we get a SPAM-free estimate of the diamond norm of the target circuit.} \label{fig:spamreference} 
\end{figure}

For the ensemble generating the ${\mathcal L}$'s we use layered circuits with brickwork, alternating, two qubit layers and single qubit layers drawn uniformly at random from the single qubit Clifford group as shown in Fig.~\ref{fig:Lcircuit}.  A number of layers logarithmic in the number of qubits is both necessary and sufficient for the scrambling we require.  That is, if we take an operator drawn uniformly at random from the Pauli group and apply some sample from ${\mathcal L}$, with high probability the resulting Pauli is indistinguishable from a second, uncorrelated random sample. A logarithmic depth is necessary because, with high probability, a random Pauli string will have a sequence of consecutive $I$'s whose length is logarithmic in $n$.  With linear connectivity (the hardest topology to scramble) we have to progressively step in from the boundary in order to modify the $I$'s in the interior of this region.  A logarithmic length circuit is also sufficient since a random Clifford circuit of log depth, even in a linear topology, generates an approximate unitary 2-design \cite{logdesgin}.  From some very coarse simulations we find that for up to $1000$ qubits $4-6$ layer deep brickwork circuits are sufficient for our purposes. 

\begin{figure}[h]
\centering
\begin{align}
\begin{array}{c}
\Qcircuit @C=1em @R=1em {
    & \gate{C}  & \ctrl{1} &
        \gate{C}  & \qw & 
        \gate{C}  & \ctrl{1} &  \qw  & \cdots&&
        \gate{C}  &  \qw &
        \gate{C}&\\
     & \gate{C}  & \control \qw&
        \gate{C}  &\ctrl{1} &  
        \gate{C}  &  \control \qw  & \qw&\cdots &&
        \gate{C} &  \ctrl{1} &
        \gate{C}&\\
     & \gate{C}  & \ctrl{1}&
        \gate{C} & \control \qw&  
            \gate{C} &\ctrl{1} &   \qw  &\cdots&&
        \gate{C}  & \control \qw &
        \gate{C}&\\
     & \gate{C}  & \control \qw&
        \gate{C} &\ctrl{1} & \gate{C} & \control \qw &  \qw  &\cdots&&
        \gate{C}  & \ctrl{1} &
        \gate{C}&\\
    & \gate{C}  & \qw &
        \gate{C} & \control \qw&
            \gate{C} &\qw  &  \qw  &\cdots&&
        \gate{C}  &\control \qw &
        \gate{C}&
        }
\end{array}\nonumber
\end{align}

\caption{Random brickwork layered circuit.  The CZs could easily be substituted with any other 2-qubit entangling Clifford gate.}    \label{fig:Lcircuit} 
\end{figure}

There has been a small amount of error assumption sleight of hand.  We need that the errors in both the SPAM and the log-depth circuit are the same for the Cliffordized and the reference circuits.  In other words, the SPAM and ${\mathcal L}$ need to have the same error properties regardless of whether ${\mathcal C}$ is present or not, which is a stronger set of assumptions than the PTA alone.

\subsection{SPAM removal with measurement mitigation}\label{sec:measmit}

The final technique we propose is the removal of SPAM errors with measurement error mitigation \cite{MeasMitigation}.  Depending on the hardware and software platforms this may be extremely easy to implement.  For example, with the IBM deployed devices in this work measurement error mitigation is turned on with a simple option flag in Qiskit's EstimatorV2 \cite{Qiskit}. 

The ease of implementation comes at the cost of assumptions. Instead of only assumptions about the physical nature of the SPAM error, we also may need to validate the measurement error mitigation technique and its implementation.  It's also worth pointing out that the techniques in Sec.~\ref{Sec:Bounds} require that the error processes are described by physical (i.e., completely positive, trace-preserving) maps.  When error mitigation is used, especially when considering imperfect model fitting, the effective error channel need not be physical. 

All this said, the reference circuit proposal in the previous section is nothing more than a specific measurement error mitigation protocol.  The assumption, therefore, is not whether to mitigate or not but whether to trust a given platform's native mitigation.  We will compare both measurement error mitigation methods in the next section.

\section{Experimental validation and benchmarks}\label{sec:exp}

In this section we will validate some of the arguments from the previous sections, using experiments on the IBM fleet of deployed heron devices -- {\it ibm\_fez,ibm\_kingston ibm\_marrakesh} and {\it ibm\_torino} -- and simulations.  We will show that the infidelities of random Cliffordizations of a non-Clifford target circuit are indeed fairly uniform, are approximately equal to the diamond norm of the target circuit, and compare our different SPAM mitigation techniques.  We will demonstrate a volumetric benchmark, \cite{volumetric}, built around random Cliffordizations and finally we will provide some comparisons to random circuit sampling, \cite{supremacy19,supremacy24}. 

For all results in this section, we will use the brickwork circuits shown in Fig.~\ref{fig:Lcircuit}. When these circuits are split apart into disjoint layers they have the same form as the direct, simultaneous randomized benchmarking circuits used in the layer fidelity protocol~\cite{layerfid} and so we concurrently run these layer circuits as a baseline comparison that gives a naturally SPAM-error-free estimate of the fidelity of each layer.  The layer fidelity protocol places stricter constraints on Markovianity, assuming that the gates in disjoint layers have the same errors independent of where they occur in the context of the brickwork circuits.

When looking back at some of the techniques in the previous sections of this manuscript, one notices that there are a lot of randomization steps:  random Pauli twirls, random Cliffordizations, possibly random SPAM references. Practically, if we want to output the fidelity averaged over a handful of Cliffordizations, all of these randomizations can be simultaneously applied. For example, by randomly sampling the single qubit Cliffords we have already implicitly performed a Pauli twirl.  Unless otherwise stated for the purpose of spot-checking our protocol we will combine all of these sampling steps and refer to the sampled output as a random Cliffordization. 

The raw data from this section is included as csv files \cite{merkel_dataset}.

\subsection{Testing the uniformity of infidelities in the PTA}\label{Sec:fidsthesame}
We now provide evidence that the infidelities of random Cliffordizations of a non-Clifford target circuit are all approximately equal, and all approximately equal to the diamond norm of the target circuit. This is the last section in which we will separate the different randomization steps, but it is important in order to show that all Cliffordizations do indeed have the same performance.

\begin{figure}
\centering
\includegraphics[width=0.48\textwidth]{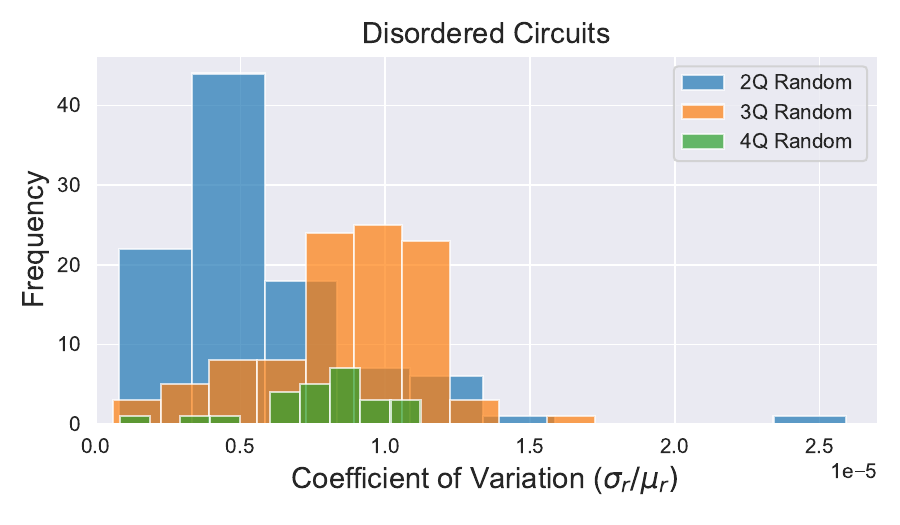}\\
\includegraphics[width=0.48\textwidth]{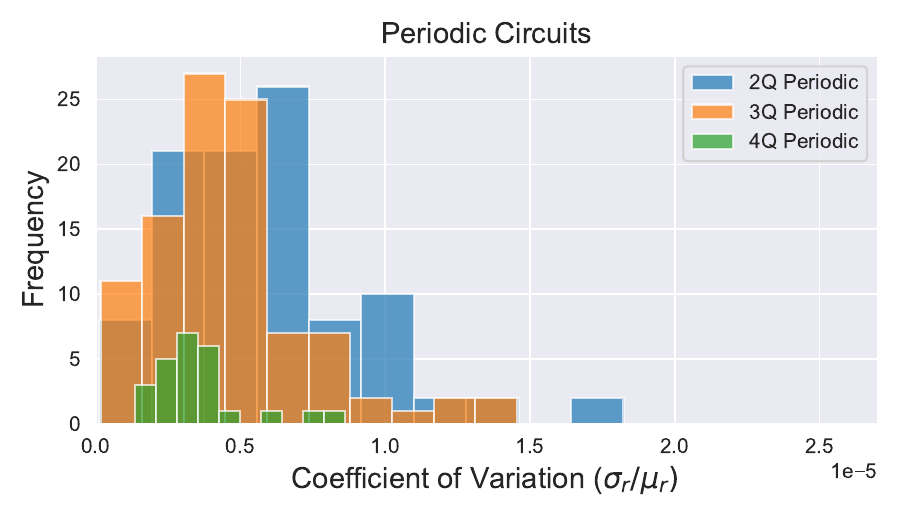}
\caption{Testing the uniformity of Cliffordized circuit infidelities. We selected $K$ $n$-qubit test circuits for $n=2$, 3, and 4 (with $K=100$ for $n=2$ and $n=3$, and $K=25$ for $n=4$) and a different Pauli stochastic error model with randomly-selected error rates for each test circuit (details in main text). We then created 500 Cliffordizations of each of those test circuits, and simulated all Cliffordizations under the error model selected for that test circuit. We did this for $K$ test circuits that are disordered (these circuits had independently-sampled random layers) and test circuits that are periodic (these circuits repeated the same pair of randomly-sampled layers many times). We computed the process infidelity $r(\cal C)$ for each Cliffordized circuit, and we computed the mean $\mu_r$ and standard deviation $\sigma_r$ of all 500 Cliffordized circuits corresponding to each test circuit. Here we show a histogram of the coefficients of variation $\sigma_r/\mu_r$ for the Cliffordizations of each test circuit. As predicted by our theory, the variation in the process infidelities $r(\cal C)$ over different Cliffordization of the same test circuit is very small.}\label{fig:simulations-1}
\end{figure}

We used numerical simulations of noisy 2, 3, and 4-qubit circuits to test whether a general circuit's diamond distance error ($d_\diamond({\mathcal U})$) is approximately equal to the process infidelities of random Cliffordization of that circuit ($r(\mathcal{C})$). Numerical simulations enable us to exactly compute $d_\diamond({\mathcal U}$) (note that the diamond norm error is difficult to measure experimentally) and $r(\mathcal{C})$ without the complications of needing to mitigate SPAM error, which is inevitable in experiments. We did not simulate circuits on more than 4 qubits because the cost of computing an $n$-qubit channel's diamond norm grows quickly with $n$ (each 4-qubit diamond norm calculation took 2 hours on an 80-core machine with 690Gb of RAM). We sampled \emph{disordered} random brickwork circuits on 2, 3, and 4-qubits (assuming ring connectivity), with various depths up to 200 pairs of layers (we sampled and simulated 100 circuits for $n=2$ and $n=3$, and 25 circuits for $n=4$). The single-qubit gates were independent samples from the Haar distribution, and each gate was implemented as a $Z(\phi_1) X_{\pi/2} Z(\phi_2) X_{\pi/2} Z(\phi_3)$ sequence. We also sampled \emph{periodic} circuits in which we repeated a single, randomly-sampled pair of circuit layers (a two-qubit gate layer and a one-qubit gate layer) many times. Each target circuit was simulated under an error model in which (i) each two-qubit gate is subject to stochastic Pauli noise with randomly sampled rates for the 15 non-identity two-qubit Pauli errors, and (ii) each $X_{\pi/2}$ gate is subject to a stochastic Pauli noise with randomly sampled rates for the 3 non-identity two-qubit Pauli error (this is a Markovian error model, i.e., each gate's errors were the same for each application of that gate, but note that we sampled a new error model for each target circuit). Each two-qubit (one-qubit) gate's total error rate was a uniformly random value between 0 and $10^{-3}$ ($10^{-4})$. For each target circuit, we also simulated 500 Cliffordized circuits under the same error model as used to simulate the target circuit. For each Cliffordized circuit, we computed $r(\mathcal{C})$ and for each target circuit we computed $d_\diamond({\mathcal U})$.

\begin{figure}
\centering
\includegraphics[width=0.48\textwidth]{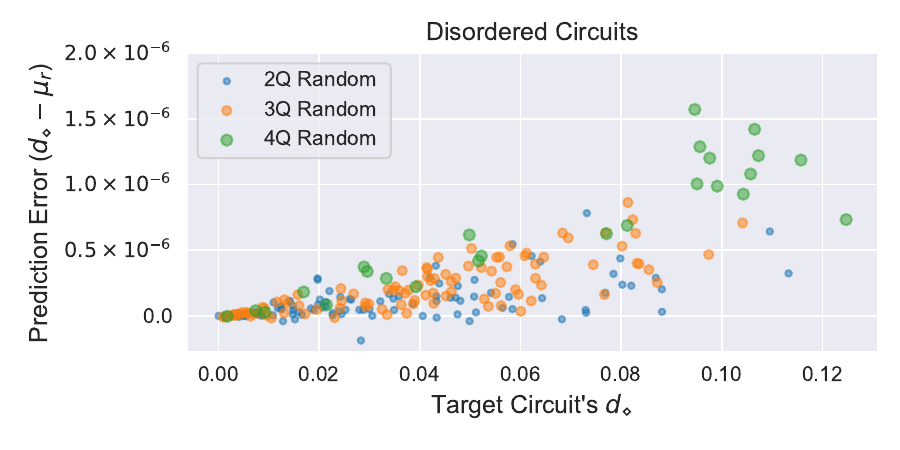}\\
\includegraphics[width=0.48\textwidth]{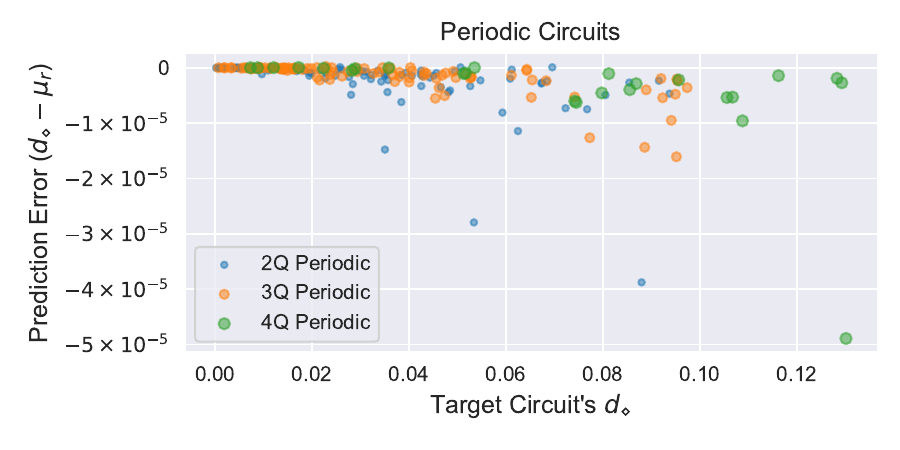}
\caption{Accuracy of diamond norm predictions from Cliffordization. We tested the accuracy with which the infidelities of Cliffordized circuits can predict the diamond distance error ($d_{\diamond}$) of a target non-Clifford circuit, using the simulated data described in Fig.~\ref{fig:simulations-1}. We find that the difference between the mean infidelity of the Cliffordized circuits $\mu_r$ and their corresponding target circut's  $d_{\diamond}$ is very small, for both classes of target circuit that we simulated: disordered circuits (top) and periodic circuits (bottom). The discrepancy increases as $d_{\diamond}$ increases, as predicted by our theory.}\label{fig:simulations-2}
\end{figure}

Figure~\ref{fig:simulations-1} shows the variation in the process infidelities of the 500 Cliffordizations of each target circuit, divided into the periodic and disordered target circuits. We computed the mean $\mu_r$ and standard deviation $\sigma_r$ of the process infidelities of the 500 Cliffordized circuits corresponding to each test circuit, and in Figure~\ref{fig:simulations-1} we show a histogram of the coefficients of variation $\sigma_r/\mu_r$ for the ensemble of Cliffordizations of each test circuit. As predicted by our theory, the variation in the process infidelities $r(\cal C)$ over different Cliffordization of the same test circuit is very small: for every test circuit it is below $2 \times 10^{-5}$. We propose using the mean of the infidelities of randomly-sampled Cliffordizations of a target circuit as a proxy for the diamond distance error $d_\diamond({\mathcal U})$ of the target circuit. Therefore, in Figure~\ref{fig:simulations-2} we compare $d_\diamond({\mathcal U})$ to $\mu_r$ for all the test circuits we simulated. We find that $|d_\diamond({\mathcal U}) - \mu_r|$ is below $2\times 10^{-6}$ for every disordered circuit we simulated, and below $5 \times 10^{-5}$ for every periodic circuit, with $d_\diamond({\mathcal U})$ ranging up to around $0.12$. This is consist with our theory.

\begin{figure}
\centering
{\bf 4 qubits and 20 layers}
\includegraphics[width=0.48\textwidth]{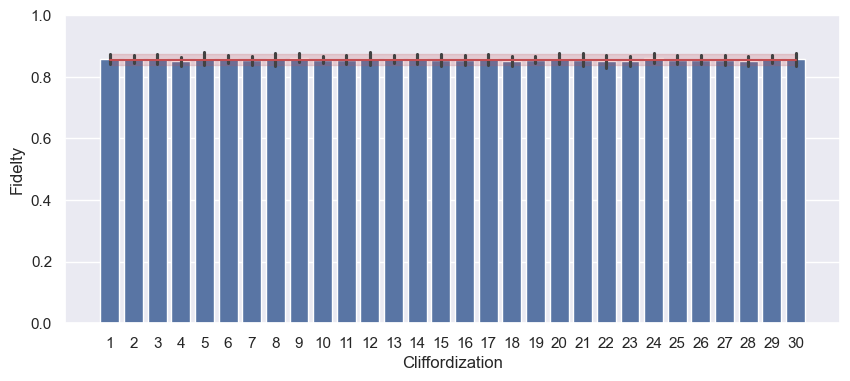}\\
{\bf 15 qubits and 20 layers}
\includegraphics[width=0.48\textwidth]{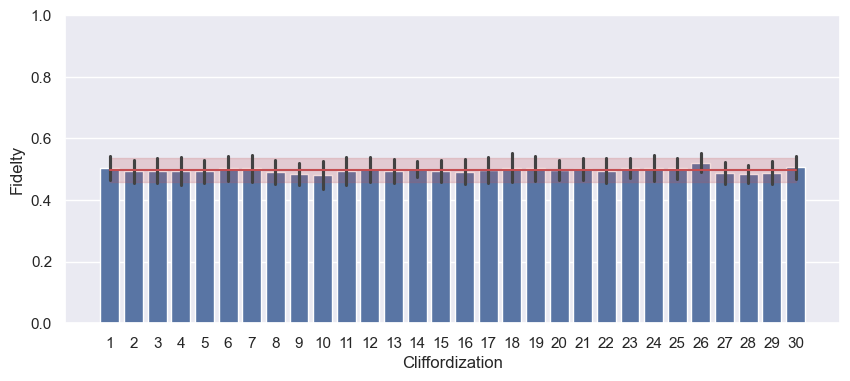}
\caption{ Testing that all random Cliffordizations have nearly the same process fidelity on {\it ibm\_fez}.  We use random brickwork circuits for 4 qubits (top) and 15 qubits (bottom) with 20 layers and plot the results from 30 different random Cliffordizations (blue columns and associated error bar).  For each Cliffordization we measure 30 independent Pauli stabilizers, and for each stabilizer we implement 32 Pauli twirls with 100 shots a piece.  SPAM was removed with measurement error mitigation.  The red line and region describe the sample mean and standard deviation which were $0.856 \pm 0.016$ and $0.497 \pm 0.039$ for the 4 and 15 qubit cases respectively.        
 }\label{fig:Exp_twirling}
\end{figure}

It is infeasible to estimate diamond distance error beyond a few qubits, in either experiment or simulation  (without strong noise model assumptions). However, we can still test the uniformity of the fidelities of Cliffordizations of a target circuit. We explored this in experiments on {\it ibm\_fez}, with 4-qubit and 15-qubit circuits. The results are shown in Fig.~\ref{fig:Exp_twirling}. The distribution of fidelities are very tightly distributed about the mean. In the absence of any other knowledge of the error model, this sort of test can build confidence in the reliability of random Cliffordizations as a benchmark for arbitrary circuits' performance.

\begin{figure}
\centering
{\bf \quad {\it ibm\_torino}}
\includegraphics[width=0.48\textwidth]{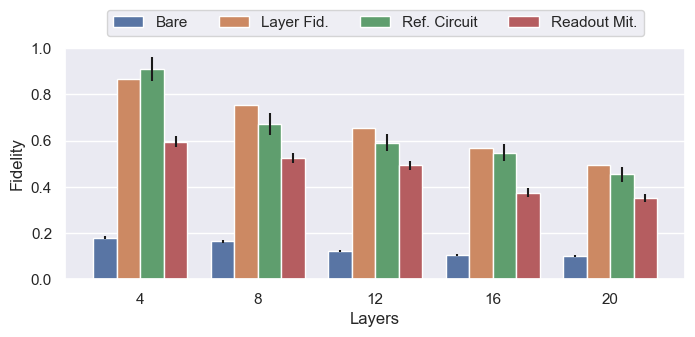}\\
{\bf \quad {\it ibm\_marrakesh}}
\includegraphics[width=0.48\textwidth]{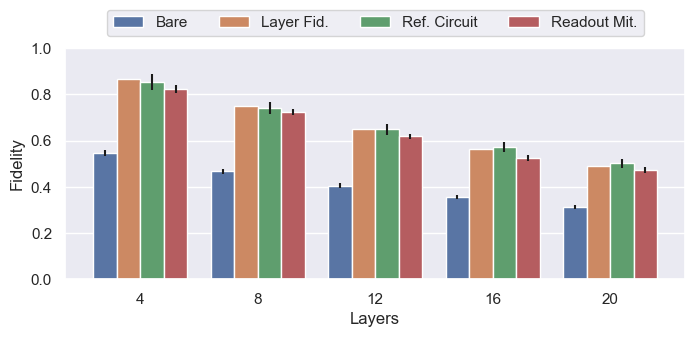}\\
{\bf \quad {\it ibm\_fez}}
\includegraphics[width=0.48\textwidth]{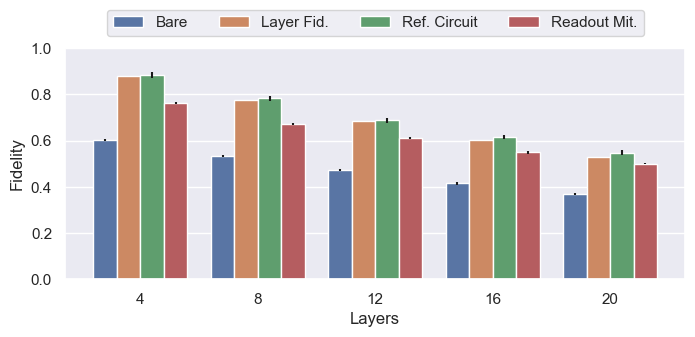}
\caption{Comparison of different measurement error mitigation methods: unmitigated (blue), a method based on layer fidelity estimates (orange), the reference circuit method from Sec.~\ref{Sec:ref_measurement} with a scrambling circuit of depth 4 (green) and measurement error mitigation using the default options in Qiskit's EstimatorV2 (red). These experiments were performed on 15 qubits with variable length brickwork circuits.  In all cases we looked at 50 random Cliffordizations with 1000 shots per randomization.  Error bars are the standard error.    
 }\label{fig:mitigation}
\end{figure}

\subsection{Comparing SPAM mitigation methods}\label{sec:SPAM_test}

We now compare the performance of the different SPAM robust implementations of direct fidelity estimation from Sec.~\ref{sec:spamrobustdfe}.  We performed 15-qubit experiments for random brickwork layered circuits of depth 4, 8, 12, 16 and 20 Fig.~\ref{fig:mitigation}.  We show results from both the reference circuit and native measurement mitigation techniques, as well as the unmitigated  data. In addition, we provide an estimate of the gate error given by the layer fidelity \cite{layerfid} that was run concurrently with the other Clifford circuits. That is, we estimate the fidelities of the two layers from independent benchmark experiments (which robustly fit with exponential decays but assume full-Markovianity) and construct an estimate from the appropriate powers of those fidelities.  

Ideally, if the errors are Markovian, there should be agreement between the layer fidelity estimate and the mitigated experiments (all but the blue, unmitigated bars).  We generally see very good alignment between the layer fidelity estimates and the reference circuit method, but for {\it ibm\_torino} (and to a lesser extent {\it ibm\_fez}) we see a separation from Qiskit's native readout mitigation.  It's also worth noting that the variance in all the mitigation methods are very dependent on the magnitude of the SPAM errors to be mitigated (here seen by the relative heights of the blue and orange bars).  This is to be expected seeing as we aren't altering the number of shots based on SPAM error estimates.  If our errors are Markovian, it would appear the reference circuit method is more accurate than Qiskit's measurement mitigation, however, the reference measurement method does have quite a bit more overhead. Perhaps the takeaway from this section is that one should test these methods against each other in order to ascertain which approach is best for a given application and required accuracy.    

\begin{figure*}
\centering
\begin{tabular}{ccc}
{\bf \quad {\it ibm\_fez}}&{\bf\quad {\it ibm\_torino}}\\
\includegraphics[width=0.4\textwidth]{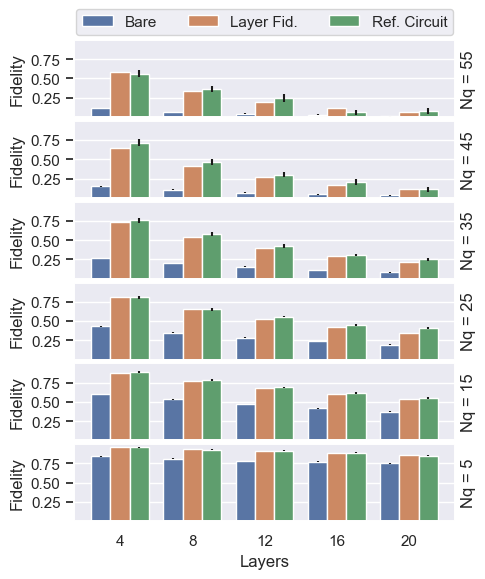}&
\includegraphics[width=0.4\textwidth]{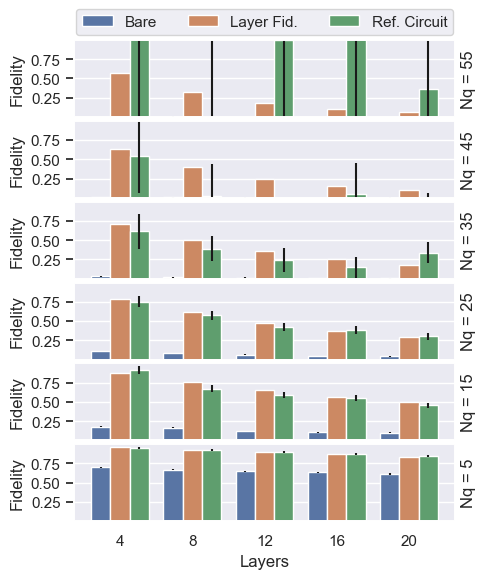}\\
{\bf \quad{\it ibm\_marrakesh}}&{\bf \quad{\it ibm\_kingston}}\\
\includegraphics[width=0.4\textwidth]{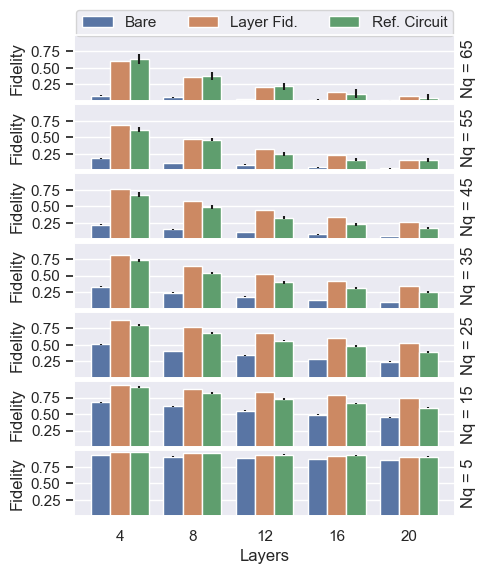}&
\includegraphics[width=0.4\textwidth]{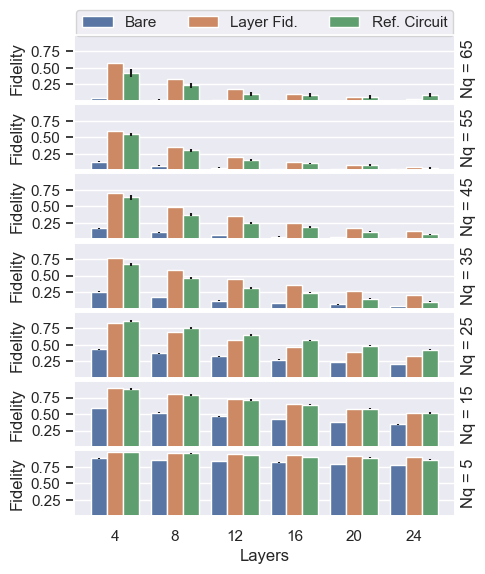}
\end{tabular}
\caption{ Volumetric plots for {\it ibm\_fez, ibm\_torino, ibm\_marrakesh} and {\it ibm\_kingston} for brickwork circuits of different widths (rows) and depths (columns).  For each circuit we run 50 randomizations with 1000 shots apiece (10,000 shots for {\it ibm\_kingston}).  The bars represent the unmitigated values (blue), the layer fidelity estimate (orange), and the reference circuit method with a scrambling circuit of depth 4 (green). Error bars describe the standard error.}\label{fig:volumetric}
\end{figure*}

\subsection{A volumetric benchmark from Cliffordization}

In this sub-section we use the Cliffordization of random brickwork circuits to create a volumetric benchmark as described in \cite{volumetric}.  We estimate the fidelity of random brickwork circuits of up to 65 qubits and 24 layers on IBM's deployed heron devices, Fig.~\ref{fig:volumetric}.  For each circuit we show the unmitigated values, an estimate from the layer fidelity and the Cliffordized estimate using the reference circuit method.  

Overall we see excellent agreement between the layer fidelity estimate and the results of Cliffordization right up until we lose all signal from the unmitigated expectation values.  For {\it ibm\_torino} we lose signal at around 25 qubits, for {\it ibm\_fez} 55 qubits and for {\it ibm\_marrakesh} and {\it ibm\_kingston} we still have signal out to 65 qubits.  This loss of signal is primarily due to SPAM, however, without increasing the number of shots/randomizations it can be hard to resolve unmitigated fidelities that are on the order of $1 \times 10^{-3}$.

These volumetric plots broadly confirm the results of the layer fidelity experiments.  It is worth noting that each of these experiments required about 10 minutes of QPU time, while layer fidelity instead took about 20s.  Both of these benchmarking tools have their uses, but for IBM's heron devices we would feel comfortable using the layer fidelity to predict the SPAM-free performance of these brickwork circuits. Having multiple metrics to test against each other is still important in order to build confidence after major hardware or software upgrades, and it still may be the case that for more exotic patterns of two qubit gates Cliffordization may show us errors that we've somehow missed with the amplification circuits used in the layer fidelity experiments. 

Before moving on, there were some subtleties in achieving agreement between layer fidelity and Cliffordization for the higher performing systems, {\it ibm\_marrakesh} and {\it ibm\_kingston}.  First, for smaller CZ errors the contribution of the single qubit gates to the overall layer fidelity is greater, therefore it is crucial to use the same single qubit Clifford decompositions for the two metrics (Qiskit natively uses a decomposition that is optimal in the number of $X_{\pi/2}$ gates whereas Cliffordization uses a fixed length decomposition).  Secondly small drifts in error rates can lead to large changes in the aggregate fidelity when considering circuits with 65 qubits and 800 entangling operations.  We found that, for our purposes, layer estimates were valid on timescales of minutes to an hour.  In Fig.~\ref{fig:volumetric}, each row (fixed qubit number) could be packaged into a single qiskit job and it was beneficial to update our estimates of layer fidelity and spam in between each job.

\subsection{Comparison to random circuit sampling}

\begin{figure}
\centering
\includegraphics[width=0.48\textwidth]{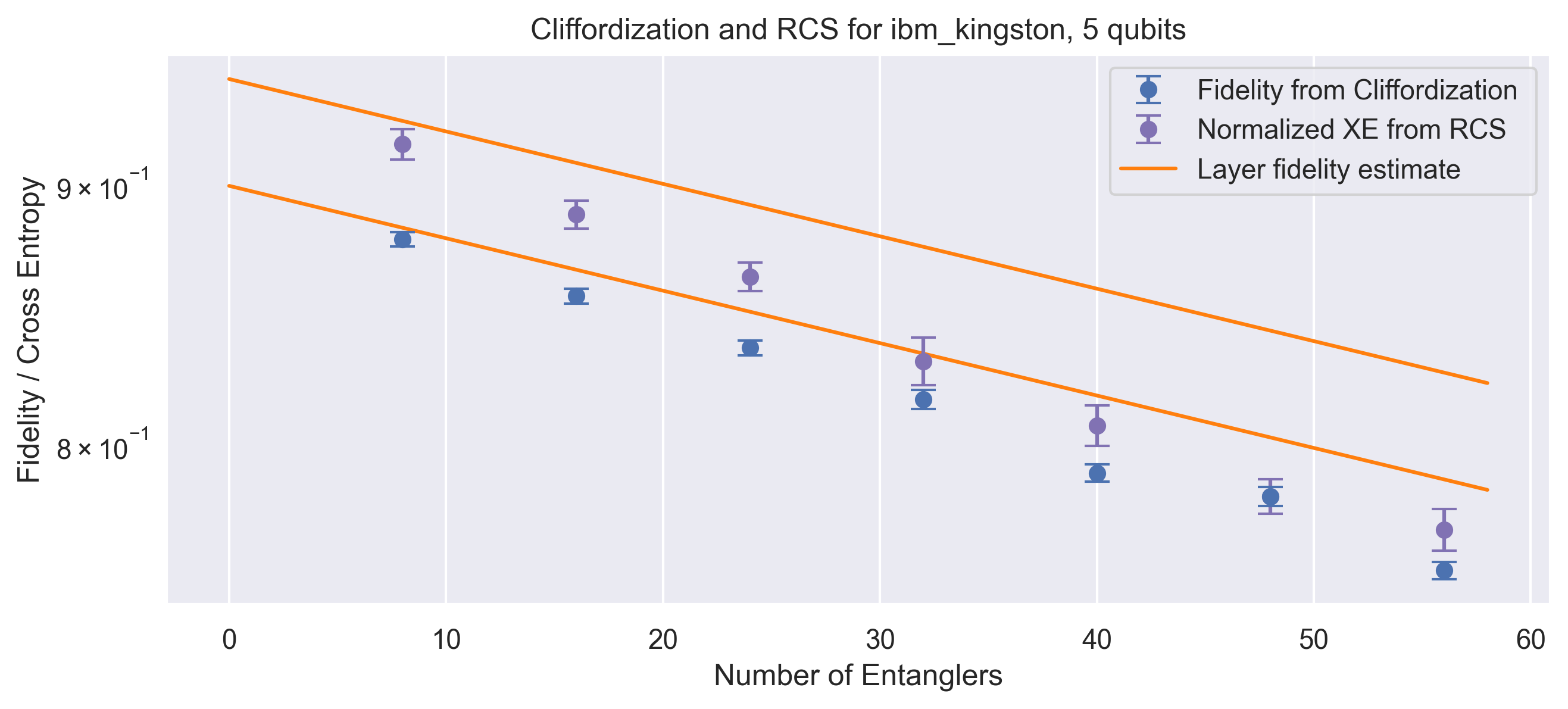}\\
\includegraphics[width=0.48\textwidth]{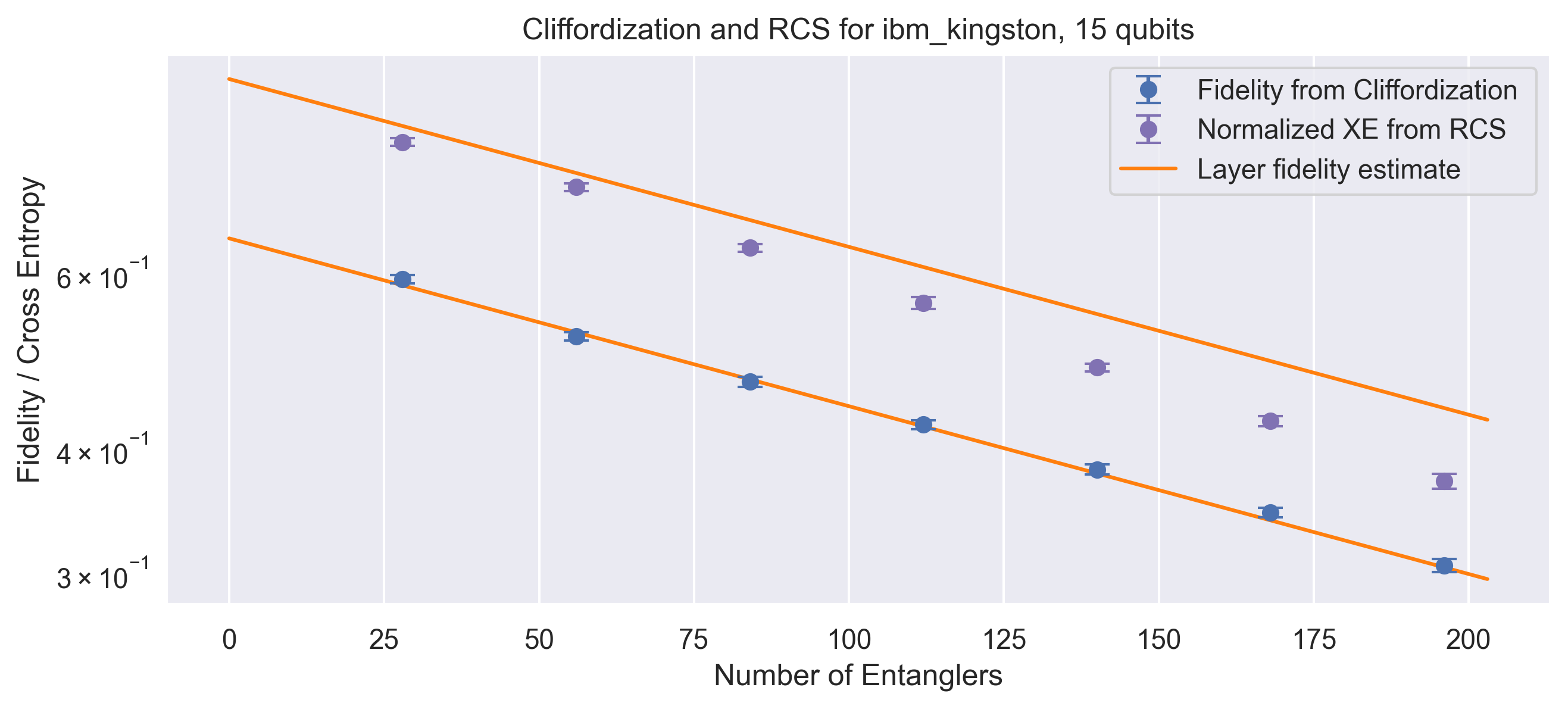}
\includegraphics[width=0.48\textwidth]{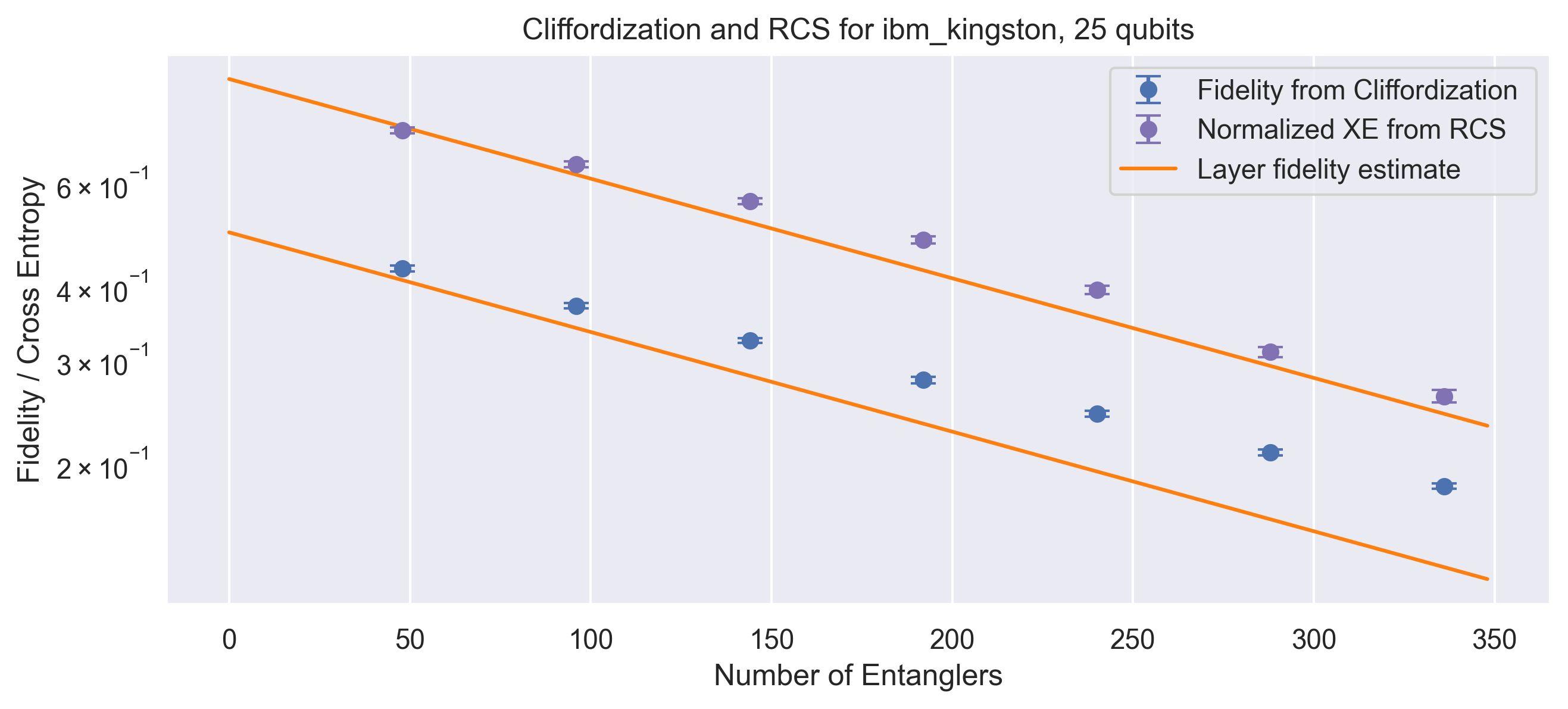}
\caption{A comparison between cliffordization (blue) and RCS [random circuit sampling] (purple) for 5 (top), 15 (middle) and 25 (bottom) qubit brickwork layer circuits on {\it ibm\_kingston}.  The orange curves are predicted values using the y-intercept of the data (log scale) as a SPAM prefactor and the slope is the layer fidelity (measured independently).}\label{fig:CliffRCS}
\end{figure}

As a final cross-validation of random Cliffordization we compare the fidelity estimates to those obtained from random circuit sampling (RCS) \cite{supremacy19,supremacy24}.  Measurements of cross-entropy are sensitive to SPAM errors, and so for this sub-section we will compare to the bare (non-SPAM robust) direct fidelity estimates from Cliffordization.  SPAM-sensitive direct fidelity estimation with high SPAM errors is outside the domain where Cliffordization yields rigorous performance bounds, however, we find the comparison between layer fidelity, Cliffordization, and RCS to be informative. 

We implement our random circuits using the same brickwork circuits as in Fig.~\ref{fig:Lcircuit}, but now we replace the random single qubit Clifford gates with gates chosen uniformly at random from the single qubit Haar measure.  It should be noted that this class of one-dimensional circuits does not satisfy some of the hardness of classical simulation criteria of the circuits in \cite{supremacy19,supremacy24} and as a result this section is meant primarily as a comparison of benchmarking techniques, not as a demonstration of quantum advantage.  Our Haar brickwork circuits have the same depths and widths as those from Fig.~\ref{fig:volumetric} for {\it ibm\_kingston} and were taken concurrently with the data in that figure.  As a technical note, these plots go out to $28$ layers rather than 24 layers since we are not removing the error contribution of $4$ layers using the reference circuit mitigation protocol.  For each size circuit we do 20 Haar randomization and take 10,000 shots.  We use Qiskit's native twirling to divide these 10,000 shots a further 20 times into different Pauli and measurement twirls.  For smaller width circuits (5,15, and 25) it is pretty straightforward to run the classical computation on a laptop, and the results are shown in Fig.~\ref{fig:CliffRCS}.  Here we've plotted the normalized, linear cross-entropy, which is defined as 
\begin{align}
    {\rm XE}(p_{\rm exp},p_{\rm ideal} )  =  \frac{2^n p_{\rm exp} \cdot p_{\rm ideal} - 1}{2^n p_{\rm ideal} \cdot p_{\rm ideal} - 1},
\end{align}
where $p_{\rm exp}$ and $p_{\rm ideal}$ describe a vector of the experimental and ideal frequencies for observing output bit-strings.  Both the fidelity and the linear cross-entropy are plotted versus the number of entangling gates (as opposed to number of layers).  

\begin{figure}
\centering
\includegraphics[width=0.48\textwidth]{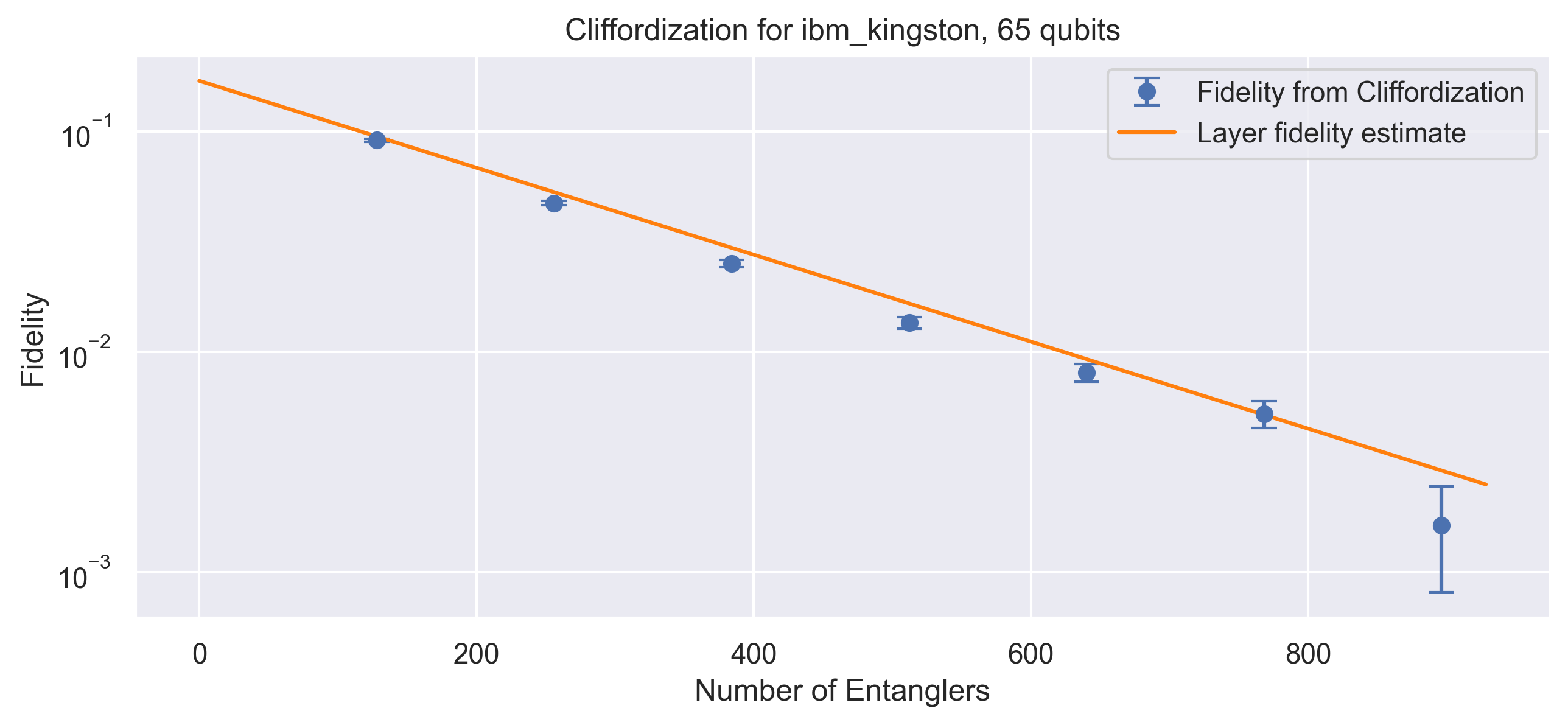}\\
\caption{Cliffordization for 65 qubits on {\it ibm\_kingston} as a function of the number of entanglers. The layer fidelity estimated curve is as described in the caption of Fig.~\ref{fig:CliffRCS}. For this data the EPLG was measured to be $3.6\times 10^{-3}$ (or a dressed Pauli error~\cite{supremacy24}/process error of $4.7\times 10^{-3}$)  Random circuit sampling data was taking concurrently and is provided online~\cite{merkel_dataset}.} \label{fig:Cliff65}
\end{figure}

We generally see decent agreement between the fidelities estimated from layer fidelity, Cliffordization, and the linear cross-entropy as measured with RCS.  There will always be an offset between a fidelity and a XE estimate due to the effects of SPAM (e.g., a bitflip at the terminus of a circuit can lead to a negative expectation value but only a zero cross-entropy). For finite depths it is also the case that RCS will typically overestimate the fidelity, as was observed  and discussed in detail in \cite{GaoRCS}.  For larger circuits it becomes prohibitively expensive to run the classical post processing to calculate the linear cross-entropy.  Additionally, for 65 qubits at our current error rates, it becomes costly in terms of the number of shots needed to resolve the fidelities from Cliffordization.  These fidelities are approaching $1 \times 10^{-3}$ which require over a million shots in order to resolve to reasonable precision.  For that reason we've repeated the 65 qubit experiment 3 times and averaged (see Fig.~\ref{fig:Cliff65}).  With 10,000 shots, 50 randomizations, and 3 runs we get a total of $1.5$ million shots per data point.  The data for Fig.~\ref{fig:Cliff65} took a little less than an hour of QPU time, which is now far more expensive than the approximately 20s of QPU time for the layer fidelity estimate. For 896 entanglers on a chain of 65 qubits of {\it ibm\_kingston} we measure a SPAM-sensitive fidelity of a Cliffordized random brickwork circuit of $1.6\pm0.8\times 10^{-3}$.  While the theory and data from Fig.~\ref{fig:CliffRCS} shows that this value is a lower bound for the cross-entropy, calculating the exact value from the data~\cite{merkel_dataset} we leave as an open problem.   

\section{Summary}

In conclusion, we have shown that all circuits with the same pattern of entangling gates have the same infidelity and diamond distance, up to small corrections, for a broad class of error models that admit Pauli twirling (the PTA).  In other words, for these error models one can bound the performance of any generic circuit by estimating the fidelity of a proxy circuit composed of only Clifford gates. We have discussed ways to estimate this fidelity in a SPAM robust manner and demonstrated these methods in simulation and experiment on IBM Quantum's heron devices.  For IBM's hardware, Cliffordization confirms the findings of the simpler layer fidelity experiment. This suggests that, Cliffordizations need only be run periodically as a spot check of layer fidelity, or alternatively to bound the performance of more exotic structured application circuits. As a benchmark Cliffordization is advantageous over RCS since the output is both verifiable and a fidelity measure.  

One outstanding area of interest that was mentioned in Sec.~\ref{sec:measmit} is the interplay between the methods in this manuscript and error mitigation.  Formally, the bounds discussed are only valid if the effective error channels are described by completely positive and trace preserving maps (which may not be the case for error mitigation with model mis-estimation).  We suspect Cliffordization can still be a useful qualitative tool for understanding performance even in the case of mitigation, but more effort is required in order to construct all but the loosest formal bounds.  Mitigation not withstanding, Cliffordization allows us to estimate both average and worst case circuit performance and therefore can be used as a predictor of the performance of any specific application circuit (i.e. accuracy of expectation values, sampling distributions, etc.).

Surely, the long-term future of quantum benchmarking will assume high accuracy and our benchmarks will look like their classical counterparts: time to solution for meaningful, verifiable, real world problems.  For near-term benchmarks of accuracy we may be forced to choose less meaningful problems, but hopefully this work argues that we need not discard verifiability as well.  Under the reasonable assumptions in this manuscript we've shown there aren't errors hiding in the dark, exponentially-sized corners of $SU(2^n)$ \footnote{Forgive the tortured metaphor. Spheres are spiky, they do not have corners} that we can't probe with Clifford circuits alone. Cliffordization allows us to fully characterize the accuracy of our hardware's performance.  If benchmarks do not themselves provide utility (e.g. random circuit sampling \cite{supremacy19,supremacy24} or quantum volume \cite{QuantumVolume}) there's no reason they shouldn't be Clifford.

\section*{Acknowledgments}
Research was sponsored in part by the Army Research Office and was accomplished under Grant Number W911NF-21-1-0002. The views and conclusions contained in this document are those of the authors and should not be interpreted as representing the official policies, either expressed or implied, of the Army Research Office or the U.S. Government. The U.S. Government is authorized to reproduce and distribute reprints for Government purposes notwithstanding any copyright notation herein. 

This material was funded in part by the U.S. Department of Energy, Office of Science, Office of Advanced Scientific Computing Research, Quantum Testbed Pathfinder Program. T.P. acknowledges support from an Office of Advanced Scientific Computing Research Early Career Award. Sandia National Laboratories is a multi-program laboratory managed and operated by National Technology and Engineering Solutions of Sandia, LLC., a wholly owned subsidiary of Honeywell International, Inc., for the U.S. Department of Energy's National Nuclear Security Administration under contract DE-NA-0003525. All statements of fact, opinion or conclusions contained herein are those of the authors and should not be construed as representing the official views or policies of the U.S. Department of Energy or the U.S. Government.

\bibliography{cliffordize}

\end{document}